\documentclass[10pt, twocolumn]{IEEEtran}

\usepackage{stfloats}
\usepackage{cite}
\usepackage{graphicx}
\usepackage{amssymb,amsmath}
\usepackage{amsfonts}
\usepackage{multirow}
\usepackage{mathrsfs}
\usepackage[usenames,dvipsnames]{xcolor}

\usepackage{algorithm}
\usepackage{algorithmic}
\usepackage{multirow}
\usepackage{amsmath}

\usepackage{graphics}
\usepackage{graphicx}
\usepackage{epsfig}
\usepackage{cases}
\usepackage{amsmath,bm}
\usepackage{wasysym}
\usepackage{subfigure}

\allowdisplaybreaks

\title{ Smart Charging for Electric Vehicles: A Survey From the Algorithmic Perspective }
\author{Qinglong Wang, Xue Liu, Jian Du and Fanxin Kong
\thanks{{The authors are with the School of Computer Science, McGill University, Montreal, Canada
 (e-mail: qinglong.wang@mail.mcgill.ca, xueliu@cs.mcgill.ca, dujianeee@gmail.com, fanxin.kong@mail.mcgill.ca).}}}
\begin{document}
\newcommand*{\QEDA}{\hfill\ensuremath{\blacksquare}}

\maketitle
\begin{abstract}
    Smart interactions among the smart grid, aggregators and EVs can bring various benefits to all parties involved, e.g., improved reliability and safety for the smart gird, increased profits for the aggregators, as well as enhanced self benefit for EV customers. This survey focus on viewing this smart interactions from an algorithmic perspective. In particular, important dominating factors for coordinated charging from three different perspectives are studied, in terms of smart grid oriented, aggregator oriented and customer oriented smart charging. Firstly, for smart grid oriented EV charging, we summarize various formulations proposed for load flattening, frequency regulation and voltage regulation, then explore the nature and substantial similarity among them. Secondly, for aggregator oriented EV charging, we categorize the algorithmic approaches proposed by research works sharing this perspective as direct and indirect coordinated control, and investigate these approaches in detail. Thirdly, for customer oriented EV charging, based on a commonly shared objective of reducing charging cost, we generalize different formulations proposed by studied research works. Moreover, various uncertainty issues, e.g., EV fleet uncertainty, electricity price uncertainty, regulation demand uncertainty, etc., have been discussed according to the three perspectives classified. At last, we discuss challenging issues that are commonly confronted during modeling the smart interactions, and outline some future research topics in this exciting area.
\end{abstract}
\begin{keywords}
Electric vehicles, smart grid, aggregator, communication networks, load flattening, regulation, uncertainty.
\end{keywords}

\section{Introduction}
\label{sec:intro}

The popularity of electric vehicles (EVs) is rising. This is attributed to the growing concerns about emissions, fossil energy depletion and city noise~\cite{chan2007state,emadi2008power}. A large-scale EV penetration has been envisioned in coming decades, e.g., the number of EVs in the world would increase by $5$ millions per year by 2020~\cite{tanaka2011technology}, while by 2014, $3.47\%$ of the U.S. automotive market is contributed by EVs~\cite{edta2015}.

The increasing adoption of EVs will also bring to power grid challenging issues. For example, the loads from charging an EV with a charging power of $19.2$kW at $80$A and $240$V, which is known as alternating current (AC) level $2$ charging standard~\cite{Yilmaz2013Review}, can be almost twenty times of that from supporting a typical North American home~\cite{ardakanian_distributed_2013}. The impacts of the significant EV charging loads will be even severer, when the charging loads are aggregated yet not coordinated. This can lead to issues including unbalance between EVs' charging demand and the power grid's power supply, more power losses and larger voltage deviation, etc. At the same time, with properly coordinated control strategies, EVs can help accelerate the realization of the smart grid, which has been envisioned to integrate EVs, renewable generations and distributed generations into the traditional power grid and utilize real-time communication to perform intelligent control strategies to coordinate bidirectional power flow~\cite{fang_smart_2012}.

To realize {both the smart grid and the smart integration of EVs}, fundamental infrastructures needs an overall upgrade.
For example, both current power grid infrastructures and EVs are not at a mature stage to handle bidirectional power flow; communication infrastructures are under development to adopt vehicular dedicated communication. 
Therefore, with more developed fundamental infrastructures, smart interactions can be realized based on real-time communication between EVs and the smart grid, {as well as via} coordinated flow control of the smart grid and intelligent control of EV charging.

\begin{figure}[t]
\centering
\epsfig{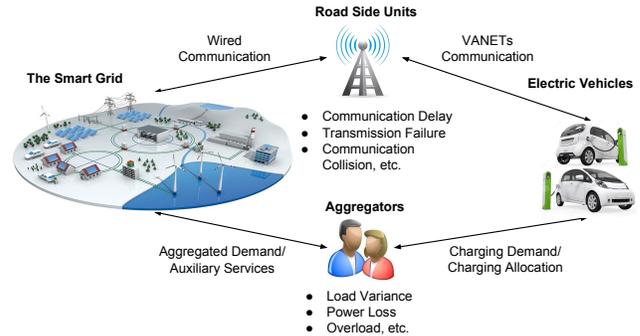}
\caption{Interaction between EVs and the smart grid. Charging loads from present EVs are aggregated by the aggregators, who represent the EV customers to negotiate with the smart grid about auxiliary services prices and electricity prices, and disseminate the charging control signals to EV customers. {For both the smart grid and aggregators, the design of control strategies needs EVs' real-time information to be collected and transmitted} to the smart grid and aggregators by road side units (RSUs) via vehicle-to-vehicle and vehicle-to-RSU (V2R) communication.}
\label{fig::smartgird}
\end{figure}


Bidirectional power flow and bidirectional communication constitute the foundation for the interaction between EVs and the smart grid, as shown in Fig.\ref{fig::smartgird}.
The power flow between the smart grid and EVs is through aggregators, who represent EV charging facilities, including power networks substations, charging stations and parking lots etc.
{Meanwhile,} the real-time information exchange between the smart grid and EVs is through communication networks, within which the access points are responsible for collecting vehicular information and disseminating control information to EVs. The power networks and communication networks together compose a complicated {network}, which needs control strategies designed to combinationally handle the interaction between EVs and the smart grid.

There have been several surveys and reviews studying the interaction between EVs and the smart grid. The potential impacts on the grid from plug-in hybrid electric vehicles (PHEVs), which is a special type of EVs, have been investigated in~\cite{green_ii_impact_2011}. An overview of EVs and fundamental industrial informatics infrastructures has been provided in~\cite{su_survey_2012}.
Both Richardson et al. \cite{richardson_electric_2013} and Mwasilu et al. \cite{mwasilu_electric_2014} have studied EVs' interaction with the smart grid with an emphasis on the presence of integrated renewable energy.
These existing surveys mainly focus on investigating the impacts of EVs to future infrastructures and the integration of renewable energy. Our work instead focuses on intensive interaction between the smart grid and EVs via aggregators. Smart interaction between EVs and the smart grid is still in its infancy and more research efforts are needed before the sought-after advantages, e.g. improved reliability and safety for the smart gird, profits for the aggregators, as well as self benefit for EV customers, can be seen in widespread industrial practice. This article aims to overview recent advances in this area. As many aspects of this interaction have been covered by existing research works, which mainly employ mathematical formulations to model this interaction, we focus on viewing the smart interaction between EVs and the smart grid from an algorithmic perspective. In particular, important dominating factors for coordinated charging from three different perspectives are studied, in terms of smart grid oriented, aggregator oriented and customer oriented smart charging. We summarize their formulations and explore the nature and substantial similarity among them. Furthermore, we discuss challenging issues that are commonly confronted during modeling the smart interaction, and outline some future research topics in this exciting area.

The rest of this survey is organized as follows. Section. II provides the background for EVs, rechargeable batteries, the smart grid and communication networks. Section. III to Section. V investigate research works about smart grid oriented, aggregator oriented and customer oriented EV smart charging, respectively. Section. VI discusses several essential open issues and points out potential future research directions. Section. VII concludes the survey.

\section{Background}\label{sec:background}

In this section, we provide not only the background information of the critical components (i.e., EVs, the smart grid, aggregators and RSUs) of the architecture shown in Fig.\ref{fig::smartgird}, but also point out their unique properties in the power networks and communication networks. These properties are crucial for future potential researches to tackle more practical issues about the interaction between EVs and the smart grid.
\subsection{Electric Vehicles}
Since the focus of this survey is to study the interaction between EVs and the smart grid, plug-in EVs (PEVs, which is capable of being charged by grid power) are mainly introduced in this section. PEVs are referred as a superset of different types of EVs with plug-in feature, including battery electric vehicles (BEVs) and PHEVs.
\subsubsection{Plug-in Electric Vehicles}
Both BEVs and PHEVs use electricity energy to drive electric motors that further provide propulsion for EVs. The electricity is supplied by on-board batteries, which in turn are charged either by the power grid when plugged in, or on-board generators when regenerative braking is enabled \cite{james2003electric}. The main difference between a BEV and a PHEV is that, a PHEV also uses fossil fuel and an internal combustion engine (ICE) to extend its driving range. These hybrid energy resources are utilized in either battery charge-sustaining (CS) mode or battery charge-depleting (CD) \cite{james2003electric}. In CS mode, fossil fuel acts as the major energy source. In CD mode, a PHEV's operations depend on electricity provided by batteries. When operating in CD mode, a PHEV can substantially reduce gasoline consumption \cite{Tuttle2012Evolution}.

Using hybrid energy sources makes PHEVs more competitive in both driving range and price than BEVs in current market. For example, a Nissan Leaf uses a $24$kWh battery pack to support a $100$-miles range. Similarly, a Mitsubishi i-MiEV can achieve the same range by using $16$ kWh battery pack. Tesla Model S is capable of providing a much larger driving range of $208$/$265$ miles with a $60$/$85$ kWh battery pack. However, the prices of EV batteries are considerably high, even though the average price of a lithium-ion (Li-ion) battery pack has been estimated to confront a yearly drop~\cite{finance2012electric}. Therefore, a BEV with a longer range usually implies a much higher price. In contrast, a PHEV can provide an equivalent driving range as an internal combustion engine vehicles (ICEVs), via using an ICE, an electric motor and a small capacity battery pack (less than $1/4$�C-$1/3$ of that equipped by a typical EV \cite{markel2010plug}). For example, a Ford Fusion can support a maximal driving range of $550$ miles. However, in term of all-electricity range, a PHEV is fairly limited by a small battery pack. For example, a Chevrolet Volt can only cover $61$ miles with a $17.1$ kWh battery pack. Meanwhile, this in turn leads to less time required for fully recharging a PHEV. For example, it only takes $3$ hours to fully recharge a Toyota Prius even using AC level $1$ charging standard~\cite{Yilmaz2013Review}. Moreover, a small capacity battery pack takes less payback time due to its lower price. According to~\cite{dickerman_new_2010}, an price of U.S. $\$0.10$ per kWh electricity is equivalent to the price of U.S. $\$0.70$ per gallon of gasoline. Further considering the government incentive, it has been estimated that a payback time only takes nearly the first quarter of the battery life.


\subsubsection{Load Characteristics and Statistical Aspects of EV Charging}
\label{subsubsec::loadchar}
It is actually the electricity carried by EVs that can influence and also benefit the smart grid. To capture the electricity usage pattern of EVs, two aspects of EV charging characteristics need carefully investigation. Firstly, as EVs moving across streets, districts and even cities, the electricity is also carried along. Hence, the spatial characteristics of EVs' mobility is inherited by the electricity stored in battery packs. Secondly, since an EV may initiate or cease its charging procedure randomly during a day, this randomness may exhibit temporal patterns when viewed at a large number of EVs \cite{Lee2012Stochastic}.

Accurate information of aforementioned aspects is crucial for smart grid planning and charging service scheduling, in order to establish more intelligent and efficient interaction between EVs and the smart grid. However, these studies require extensive data processing of EV trips. Especially when complex EV model is employed, infeasible computation can be incurred. Therefore, these characteristics needs to be investigated through statistical approaches \cite{Lee2012Stochastic,Vagropoulos2013Optimal}.

Spatial characteristics of EVs' energy usage are related to real world driving conditions, including physical conditions of road segments, traffic congestion status and driving patterns, etc. Among these, studies about EV driving pattern have been more intensely developed. Two federal driving schedules (i.e., Urban Dynamometer Driving Schedule and  Highway Fuel Economy Driving Schedule) have been widely used for analyzing PHEVs' driving pattern \cite{lohse2009drive,duoba2009calculating}. However, these models suffer from having limited duration and restricted acceleration/deceleration rates \cite{Tara2010Battery}, assuming constant electricity usage per hour per mile \cite{Adornato2009Characterizing}, and lacking considering the influence of driving distances \cite{Lee2012Stochastic}. In \cite{Adornato2009Characterizing,Lee2010Synthesis,Lee2012Stochastic}, a naturalistic driving cycles model generates significantly different instantaneous load compared with two aforementioned schedules. Specially, \cite{Lee2012Stochastic} uses a naturalistic driving cycle to study the distance distribution of a large number of EVs. 

Temporal characteristics of EV charging include the initiating and ceasing time of a charging process, and the energy usage varying over time. Lee et al. propose a stochastic model to explore the relation between the distribution of EV arrival time and departure time in \cite{Lee2012Stochastic}. The distribution of arrival time is modeled as a Gaussian distribution, conditioning on the departure time distribution (chi-square distribution). This approach captures departure and arrival time distribution via using a small number of simulations, thus mitigating the computational burden.

Despite the importance of exploring spatio-temporal characteristics of EV charging, there are few real world data sets available currently. To conquer this, a synthesized data set has been developed in \cite{Akhavan2014Developing}. The authors combine real world taxi trajectories with the features of four dominating PHEV brands in the North American market to develop a data set which contains spatio-temporal characteristics of EV charging load.

\subsection{Properties of Rechargeable Batteries}
As we have discussed the spatio-temporal characteristics of EV charging in Section. \ref{subsubsec::loadchar}, in this part, we focus on the carrier of electricity, i.e., rechargeable batteries. Li-ion batteries are prevalent in current EV market. This is because Li-ion batteries have many excellent properties, e.g., high energy density, slow self-discharge and less environmental influence, etc. These properties also make Li-ion batteries popular for many portable electronics. Charging a Li-ion battery requires very delicate control of voltage and current output of the charger. A large voltage and current fluctuation may cause destructive damages to a Li-ion battery. {Therefore, problems about charging EV batteries play a critical role for enabling the interaction between EVs and the smart grid, as well as maintaining EV batteries' lifespan.}

In the subsequent part, several unique properties of rechargeable batteries are pointed out. These properties are the key insights for differentiating charging an EV from refuelling a traditional ICEV, thus should be carefully studied.
%
%
%
\subsubsection{Charging Power Controllability}
Many research works have been conducted through utilizing continuously controllable charging power to achieve objectives including flattening loads~\cite{gan2011optimal,gan_stochastic_2012,ma_decentralized_2010}, minimizing electricity cost~\cite{ahn_optimal_2011,he_optimal_2012,ma_decentralized_2010}, maximizing overall welfare~\cite{ardakanian_realtime_2012,ardakanian_distributed_2013,ardakanian_real-time_2014,asr_network_2013,rahbari-asr_network_2013,rahbari-asr_cooperative_2014}, etc.

Currently, EVs can be charged by AC level $1$ (expected power varies from $1.4$kW to $1.9$kW, charging time varies from $4$ to $36$ hours), AC level $2$ (expected power varies from $4$kW to $19.2$kW, charging time varies from $1$ to $6$ hours) and DC fast charging (expected power can be $50$kW or $100$kW, charging time varies from $0.2$ to $1$ hour) three charging standards \cite{Yilmaz2013Review}. AC level $1$ is commonly used by on-board charger without the need for additional charging equipment installation. AC level $2$, however, requires installation for home charging and public charging. Both AC level $1$ and AC level $2$ share the same connector of SAE J$1772$ \cite{Tuttle2012Evolution}. DC fast charging is more suitable for fast charging need. The measurement and estimation of these different levels of charging loads are critical for an aggregator to achieve the economic goal while maintaining the stability and reliability of the smart grid.

However, none of the three current popular charging standards are capable of providing continuously controllable charging power~\cite{Yilmaz2013Review}. Moreover, according to~\cite{james2003electric}, charging an EV's battery packs requires very delicate control of the supplied charging power. The injected DC voltage and current must be well smoothed. Hence, without more advanced battery technique, the charging power can only vary within a {discrete set of nearly constant values.}

\subsubsection{Battery Charging Rate}
\begin{figure}[t]
\centering
\epsfig{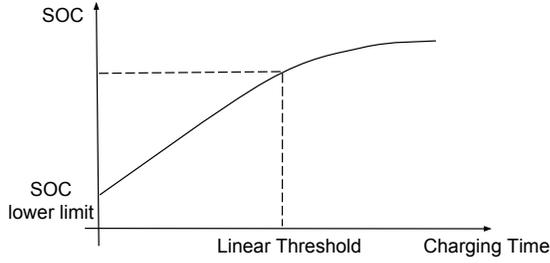}
\caption{Battery SOC obtained versus charging time spent of a Li-ion battery \cite{inventory}. The (typically) last $1/3$ of the charging curve indicates that the available battery SOC has a nonlinear dependence on the charging time spent~\cite{paryani2014fast}.}
\label{fig::ChargingCurve}
\end{figure}

Another important aspect of charging an EV is the nonlinearity between the charging time spent and the state of charge (SOC) obtained,
as shown in Fig.~\ref{fig::ChargingCurve}. The time taken for completing the final part of the charging (with higher SOC) is usually longer compared with that for the initial stage (with lower SOC)~\cite{millner2010innovative,serrao2011optimal}.
According to Fig. \ref{fig::ChargingCurve}, typically during the last $1/3$ of the charging circle, the effective charging current $I$ keeps dropping as the battery cell open circuit voltage $V_{\textrm{open}}$ keeps increasing. Thus the SOC increases nonlinearly along with more charging time spent. However, when comes to discharge, the loss of charge is linearly dependent on discharging time~\cite{paryani2014fast}.
{This inconsistency between charging and discharging is critical to differentiate EV charging tasks with different initial SOCs and equivalent relative charging demand.} It can also motivate EV customers to pursue for more efficient charging schedules. Unfortunately, this inconsistency between charging and discharging has been rarely explored yet.

\subsubsection{Battery Aging}
\label{subsubsec:battery_aging}
While charging are usually modeled as the product of electricity price and the amount of electricity drawn, another type of cost due to battery aging is more difficult to model. Fast battery aging will lower the payback of an EV customer, since the cost of Li-ion batteries comprises a significant percentage of today's EVs' prices. Therefore, battery aging is an important factor that should be taken into consideration when design coordinated charging.

Battery life is commonly regarded as the minimum between \textit{Calendar Life} and \textit{Cycle Life}. The former is the elapsed time before a battery becomes unusable. The latter is the number of complete battery charge-discharge cycles before its nominal capacity falls below certain threshold. The factor mainly influences calendar life is the temperature of the battery. While for cycle life, factors including usage pattern of batteries, battery current and the cycle depth of each usage contribute to the aging of cycle life \cite{Guenther2013604}. All mentioned factors lead to effects including loss of active mass, decomposition of electrolyte and electrodes, etc. \cite{Dubarry2012204,Lorenzo2011Optimal}, which further cause capacity fade and internal resistance increase.

Capacity fade happens each time for a charge-discharge cycle \cite{Guenther2013604}. Therefore, the lifetime of a Li-ion battery will be greatly shorten due to frequent charging and discharging, which is likely to happen when coordinated charging is designed without considering battery aging factors. An experiment result of \cite{Guenther2013604} demonstrates that battery life can be shortened due to additional cycling resulting from providing auxiliary service (peak-shaving). Therefore, despite the thriving research development regarding utilizing EVs to provide auxiliary services (frequency regulation, voltage regulation, etc.), it is crucial to consider the resulting battery aging effect. Many research works have taken battery aging into consideration. For example, in \cite{Lunz2011Optimizing}, cyclic aging cost for a half cycle has been considered when the profit of trading energy in a vehicle-to-grid (V2G) market is optimized, while in \cite{Lorenzo2011Optimal}, the cost due to fraction of battery life depleted is co-optimized. Moreover, it is also shown in \cite{Guenther2013604} that battery life can be extended if the depth of charging-discharging cycles is reduced. This is because battery aging per cycle has nonlinear dependency on the depth of each cycle.

Although there have been many research works targeting on modeling battery aging \cite{Wenzl2005373,Rong2006analytical,Smith2010Model}, however, as stated in \cite{Lorenzo2011Optimal}, the highly dynamic situations under which batteries operate rarely match the laboratory conditions used by batteries manufactures, thus may lead to inaccurate estimation of the battery aging. Moreover, to make the case even more complex, different types of Li-ion batteries have different aging characteristics, thus need different aging mechanisms \cite{Han201438,Dubarry2012204}. Additionally, according to \cite{Guenther2013604}, battery aging is also influenced by different operating scenarios, including different daily driving cycles, different charging initiated time, charging charging strategies, etc.

\subsection{EVs Interacting with Smart Grid}
The smart grid is a network of electrical components used to supply, transmit and consume electric power.
The smart grid enables bidirectional flows of energy that leads to an array of new functionalities and applications, with the aid of coordinated control, EV aggregation, V2G realization and two-way communication, etc. In the subsequent, we introduce current practice regarding aforementioned aspects, which also outlines future research directions.
\subsubsection{Coordinated EV Charging}
\label{subsubsec::coordination}

According to \cite{schuller2013electric}, EV charging coordination is categorized into (1) centralized, (2) hierarchical and (3) decentralized coordination. Centralized coordination depicts a simple architecture, where the central controller has direct control over all participated EVs. This approach, however, is less practial due to the requirement of accurate acquisition of EV status data and poor scalability \cite{li2005price}. Additionally, current power market has no support for direct control contract, as the minimum power capacity threshold is considerably higher than that of an individual EV \cite{kempton2005vehicle}. While for decentralized coordination, EV customers are assumed to have complete control over their EVs and interact with the smart grid individually through price-based mechanisms. Besides aforementioned limit of power capacity threshold, this approach may also suffer from communication overhead. Hierarchical coordination is regarded as a hybrid paradigms of both centralized and decentralized coordination. It commonly assumes the existence of an aggregator in a price-based mechanism, which operates as the intermediate between the smart grid and EV customers (as shown in Fig. \ref{fig::smartgird}). Hierarchical coordination is regarded as the most promising option for near-term realization of EV coordinated charging by both academia and industry.

By now, EV coordinated charging is more at a conceptual level, its realization requires an interweaved practice of communication standardization and the design of physical interfaces, contract frameworks as well as control algorithms. A commonly assumed price-based coordination approach, time-of-use (TOU, with low electricity price at off-peak hour and high price at peak hour) has been studied by many research works and implemented by many experiments \cite{Celebi2007Model,Celebi2012Time}, as well as utility companies (e.g., Illinois Power Company \cite{allcott2009real}, PG\&G \cite{pge}, Hydro One \cite{hydrone}, ComEd \cite{comed}, etc.).


\subsubsection{EV Aggregation}
As previously discussed, hierarchical coordination is more promising compared with centralized and decentralized coordination. To realize hierarchical coordination, a fundamental requirement is a proper EV aggregation scheme. Actually, there have already been some practical EV aggregation examples in some countries. For example, MOBI.E (an industrial network in Portugal) \cite{Bessa2012Optimized}, Better Place (a company developing battery charging techniques) \cite{Andersen20092481} and the Western Danish power system \cite{Pillai2011Integration} have utilized aggregators to operate as the intermediate for energy trading, without directly control the charging rates of participated EVs. Meanwhile, more academia and industry practice of aggregation management have been investigated in \cite{Bessa2012Economic}.
\subsubsection{V2G Systems}
Hierarchical coordination can be further enhanced by providing bi-directional power flow between the smart grid and EVs, referred to as V2G. While there have been intensive studies and experiments on V2G framework and mechanism design \cite{brooks2002vehicle,kempton2008test,Pillai2011Integration}, the practice of providing prototype V2G-enabled vehicles in industry is also under rapid development \cite{ACpropulsion,nrel,Delaware}. Especially, a PHEV charger system which is capable of providing reactive power support to the grid has been tested in \cite{Kisacikoglu2010Examination}. The test result indicates that EV customers can benefit from using EVs to provide voltage regulation without causing severe degradation to the batteries (which is severer for frequency regulation when frequent charging and discharging is conducted). However, despite the technical advancement has been achieved, social response to V2G indicates high reluctance from EV customers to hand over the control right of their EVs to third parties, despite the economical attractiveness shown in \cite{Tomi2007459,Parsons2014313}.

\subsection{EVs Communication with the Smart Grid }

Besides high-accuracy and time-synchronized measurements of both power information and phase information of the power networks \cite{PMU-JD}, as well as coordinated control methods previously introduced, the aggregators also need real-time, high-precision and low-latency communication networks in order to develop effective optimal electricity dispatch decisions. Meanwhile, this communication networks is also essential for further transmission of the dispatch decisions and real-time prices back to both the generators and utilities, including microgrids, renewable generations, distributed generation, EVs, etc.
\subsubsection{Vehicular Ad-hoc Networks}
To realize mobile real-time information exchange, techniques like cellular and Wi-Fi have been employed in many research works~\cite{lee_mobile_2010,herrera_evaluation_2010}. However, none of these systems are exclusively designed for vehicular data transmission. Therefore, with less suitable communication methods, issues including inaccurate location measurement, data collision (due to high density of vehicles) and transmission failure (due to insufficient coverage) are usually inevitable due to the high mobilities of vehicles.

The vehicular ad-hoc networks (VANETs) is dedicatedly designed for vehicular information exchange~\cite{wang_mobility-aware_2014}. VANETs enables multi-hop alike communication among vehicles and road-side units (RSUs), including short-range vehicle-to-vehicle (V2V) and vehicle-to-RSU (V2R) communication.
Both the economic and time efficiency of vehicular information collection can be improved by VANETs~\cite{cheng_infotainment_2011}. Moreover, with proper deployment of RSUs, not only the real-time information transmission range can be expanded, but also the real-time transmission can be better protected from transmission failure. Therefore, VANETs are promising to be adopted by the fundamental communication networks to support the real-time information exchange between EVs and the smart grid.

\subsubsection{Communication Protocols and Standards}
The information exchange via V2V and V2R within VANETs can be enabled by dedicated short range communication (DSRC) protocol~\cite{kenney_dedicated_2011}. Based on DSRC protocol, vehicular information can be transmitted from EVs to RSUs in a multi-hop manner. The collected vehicular information can be further sent to the smart grid through wired communication (e.g., optical fiber) to lower the communication outage and delay. After receiving the aggregated information, the smart grid disseminates the control decisions to each EV through RSUs using DSRC.


Although VANETs combined with DSRC protocol can greatly improve the accuracy, transmission range and transmission reliability for vehicular information collection, communication delay will inevitably compromise these benefits. For example, delayed information collection may jeopardize the stability of the smart grid, especially when the smart grid is already operating near full capacity. Furthermore, a delayed control information dissemination can lead to unexpected driving cost. However, researches about evaluating the impacts of the transmission delay on both the smart grid and EVs remain valuable yet open.

Meanwhile, there are also many other communication protocols that have been intensively studied to be utilized in vehicular communication networks. For example, U.S. National Institute for Standards and Technology (NIST) have recognized ZigBee and ZigBee Smart Energy Profile (SEP) as the most suitable communication standards for smart
grid residential networks \cite{Peizhong2011Developing}. Meanwhile, RF mesh-based systems have been widely adopted in North America. For example, utility company PG\&E has adopted RF mesh-based system in their SmartMeter system \cite{Gungor2011Smart}. There are also other utility companies, e.g., National Grid (U.S. utility company) and SPAusNet(Australian distribution company) select WiMAX technology for building their dedicated wireless communication networks \cite{Gungor2011Smart}. Another thread of research works concerning wireless interference and congestion, due to the utilization of the same frequency bands. Hence, cognitive radio technique has been developed in \cite{Ghassemi2010Cognitive,Rong2011Cognitive,ranganathan2011cognitive} for spectrum sharing across smart grid communication networks.

\subsubsection{Communication Requirement for Smart Charging}
To enable smart grid, the communication requirement as well as related communication techniques have been intensively studied by several survey papers \cite{Kayastha2014Smart,Gungor2011Smart,Ancillotti20131665}. Therefore, in the subsequent part, we mainly discuss the fundamental communication requirements with respect to EV charging.


One promising communication hierarchy is depicted as the smart grid communicates with aggregators, and an aggregator communicates with EV customers (Section. II-C). In this hierarchy, an EV being charged can generally be regarded as a common electric appliance when the charging operation is carried out at home environment. However, recall that the charging power required for charging a single EV (Section. \ref{sec:intro}) is considerably large, EV charging needs more concerns from the smart grid. The concerns further grow when many EVs are aggregated by charging stations or utility companies (aggregators), from where the aggregated charging demand can bring severe impact to the power grid if not coordinated properly. Moreover, as EVs are expected to be capable of providing V2G service, this makes EVs also function as distributed energy sources which maintain a bidirectional power flow between EVs and the smart grid. Additionally, an EV customer's charging pattern is indicative of this customer's mobility pattern, since people are more likely to charge their EVs at home when resting, or at charging station when driving to work. Adding up all these concerns, EV charging raises more rigid communication requirements for the communications between smart grid and aggregator, and the communications between aggregators and EV customers.

\textbf{Requirements for communications between smart grid and aggregator}: the information exchange between smart grid and aggregators mainly serves for the purpose of maintaining grid stability and safety. Therefore, the requirements are more specified to this very purpose, thus including high reliability ,availability and quality-of-service (QoS) requirements. The requirement of high reliability and availability is needless to say, while the QoS requirements are required to measure the impact of delay and outage to the aggregators, for whom the main benefit originates from further providing reliable services to EV customers. QoS support in smart grid communication is still an open issue and have attracted a great amount of research works developing QoS-based framework and extended protocols \cite{Ancillotti20131665}, it is out of the scope of this work to present detailed discussion of these solutions. Note that the communications between aggregators and the smart grid are in the order of typically 15 minutes \cite{Bayram2014survey}, hence the communication latency in case of when price information is exchanged, is less critical.

\textbf{Requirements for communications between aggregator to EV customers}: as aggregators collect and maintain the charging demand from EV customers, these charging records can reveal private mobility patterns of EV customers, hence it is crucial to establish security mechanisms to provide authorized access control and protect the integrity and confidentiality of personal data. Meanwhile, besides the privacy threat of unauthorized exposure of personal data, it is also crucial to protect the smart grid from cyber attacks, since the smart grid are greatly relied on communication networks, which has been the target of various cyber attacks constantly. Two types of cyber attacks recently draw the attention from academia, i.e. load altering attack \cite{Amini2015Dynamic} and rate alteration attack \cite{Mishra2015Rate}. Besides high security concern, the communications between EV customers and aggregators usually involve information exchange for monitoring, billing and authorization, etc., hence the tolerable latencies are commonly in the scale of seconds or even milliseconds \cite{Bayram2014survey}. Additionally, since the communications are established for end users to serve for their various purpose, it is desirable for the communication being capable of differentiating services provided.

\section{Smart Grid Oriented EV Smart Charging}\label{sec:load}
Severe unbalance between EVs' charging demand and the power networks' power supply would occur if the interaction between EVs and the smart grid is uncontrolled or non-coordinated. For example, Maitra et al.~\cite{taylor_evaluations_2010} investigate the impact of PHEVs on the local distribution system of Hydro-Quebec. Their evaluation result indicates that there is a positive linear dependence between the impacts and the penetration levels of PHEVs. Moreover, uncontrolled loads can increase the temperatures of the transformers thus shorten their lifespan~\cite{roe_power_2008}. Other issues including larger voltage variance, more power losses, lower energy efficiency, larger frequency deviation and voltage deviation also jeopardize the stability and reliability of the smart grid.

However, if the interaction is intelligently controlled and coordinated, a \textit{`valley-filling'}  strategy can further mitigate the unbalance between the supply and demand by shifting large power demands to the valley of overall load profile. In this way, balanced supply and demand of active power help maintain stable gird frequency. Additionally, if the generation and consumption of reactive power are also balanced, large voltage deviation can be prevented. Furthermore, the transformer loss of life (due to increased transformer temperatures) can also be compensated by better maintenance of transformer bushings (due to more flattened loads)~\cite{Farmer2010Modeling}. The realization of these benefits can be attributed to both the time- and power-dimensional flexibility of EV charging.

\begin{figure}[t]
\centering
\epsfig{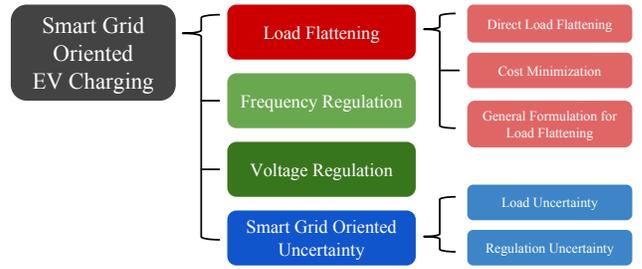}
\caption{Smart grid oriented EV charging.}
\label{fig::grid_orient}
\end{figure}

In this section, we mainly investigate research works aiming at utilizing EV charging for (1) flattening charging load, (2) providing frequency regulation service and (3) voltage regulation service, as shown in Fig. \ref{fig::grid_orient}. Most of these works stand from a grid oriented perspective to coordinate EV charging.

\subsection{Load Flattening}
\label{subsec::flattening}
In the subsequent, we split coordinated control strategies for realizing EV charging load flattening into (1)~direct load flattening and (2)~indirect load flattening. In the first case, the smart grid operator directly manages the charging loads of EVs, in order to alleviate the unbalance between EVs' charging demand and the power networks' power supply. In the second case, the smart grid operator concerns more about reducing the cost (energy generation cost and consumption cost) caused by EV charging. These problems are further shown to be equivalent to issues including improving load factor and reducing power losses. Then we present and analyse a general formulation for load flattening problems. Finally, we briefly address V2G problems with a minor modification of the general formulation.
\subsubsection{Direct Load Flattening}
\label{subsubsec::direct_control}
Many research works leverage optimization based approaches to flatten the EV charging loads.
They are generally coincident about two most common and critical assumptions: (1) the charging power $p(n,t)$ for $n$th EV at time $t$ is continuously controllable; and (2) they have perfect exogenous information $S$. The first assumption has been discussed in Section II. For the second assumption, exogenous information $S$ consists of $(N(t), s(n))$, where $N(t)$ denotes the number of EVs being charged at time $t$, $s(n)$ is a set of specific features to describe the charging demand from $n$th EV. Here we set $s(n) = \Big \{E(n,0), E(n,T), P_{min}(n,t), P_{max}(n,t) \Big \}$, where $E(n,0)$, $E(n,T)$, $P_{min}(n,t)$ and $P_{max}(n,t)$ stand for the initial stored energy, maximal requested energy at time $T$, minimal charging power and maximal charging power at time $t$, respectively. These notations will be used throughout the rest of this survey.

The research efforts falling into this category target at employing optimization based approaches to flatten the load fluctuation caused by uncertain EV charging demands, while managing to fulfil these charging demands within a certain time horizon. The common optimization objectives consist of minimizing the aggregated loads, minimizing load variances, minimizing energy costs for charging EVs, etc. {A general optimization problem for minimizing the aggregated loads over a certain time horizon can be formulated as follow:}

\begin{subequations}
\begin{align}
\mathbf{Case\:1}: \; & \underset{p}{\text{min}}
& & \sum_{t=1}^{T}\Big [ P_{B}(t) + \sum_{i=1}^{N} p(n,t) \Big ]^{2}, \label{eq:obj 1}\\
& \text{s.t.}
& &\sum_{t=1}^{T} p(n,t)\Delta_{t} \leq E(n,T) - E(n,0), \label{eq:constraint 1} \\
& & & P_{min}(n,t) \leq p(n,t)\leq P_{max}(n,t), \label{eq:constraint 2} \\
& & & n = 1, \ldots, N,\quad t = 1,\ldots,T, \nonumber
\end{align}
\end{subequations}
where $\Delta_{t}$ is the time interval of a single time slot. $P_{B}(t)$ is the base loads at $t$th time slot. In this formulation, the charging power $p(n,t)$ for each EV is continuously controllable within a specified range as shown in (2c). Meanwhile, since the exogenous information {(i.e., future base loads $P_{B}(t)$, number of EVs $N$ and their homogeneous charging interval from $t = 1$ to $T$)} is assumed to be perfectly known in advance,  an off-line approach can be employed.

{As show in (\ref{eq:obj 1})}, there are coupled decision variables in the objective, a decentralized approach can alleviate the computation burdens on the aggregators, while avoiding requiring all exogenous information from all participating EVs. Gan et al. solve this problem with gradient projection in~\cite{gan2011optimal}. It can be seen that the objective is convex with respect to the {aggregated loads $\sum_{i=1}^{N} p(n,t)$ from all participating EVs. Thus there exists a global optimum value for the aggregated loads.} Since this optimal value of aggregated loads is comprised of the combinations of individual load, multiple choices of the combination is possible (i.e., the optimal solution set is a equivalence class).

{Ref. \cite{gan2011optimal}} has been further extended in~\cite{gan_stochastic_2012}, through restricting the charging power $p(n,t)$ to be non-continuously controllable (i.e., $p(n,t)$ for each EV $n$ is fixed to $p(n)$ once initiated). The corresponding constraint is as follows:
\begin{equation}
\begin{aligned}
p(n,t)\! = \!\left\{\begin{matrix}
p(n), & \text{if EV}\, n \, \text{is charged at time}\, t,\\
0, & \text{otherwise}.
\end{matrix}\right.
\end{aligned}
\end{equation}

This optimization problem is more challenging since the decision set is discrete. The decision variable is interpreted as the {start and resume time for charging an EV with its fixed charging power.
Furthermore, according to \cite{gan_stochastic_2012}, frequently interrupted EV charging can lead to battery degradation.} Thus, the {objective is to design} the optimal charging profiles for all present EVs, i.e.,  to decide when to initiate charging. Once initiated, the charging procedure cannot be interrupted. More specifically, the aggregator has to deliver $E(n,T) - E(n,0)$ amount of energy to EV $n$ within the whole time horizon.

In order to solve this problem, Gan et al. introduce a probability distribution for all potential feasible charging profiles in~\cite{gan_stochastic_2012}. This modification can relax the decision set from discrete set to its convex hull. Thus, each EV can pick a charging profile randomly according to a updated probability distribution (obtained through solving optimization problems in a decentralized manner) within each iteration. Then each EV can generate a sequence of the random charging profiles over the iterations. Consequently, the aggregated loads of all EVs over all time slots is also a sequence of random variables. This sequence of aggregated loads and the set of aggregated charging profiles comprise a supermartingale pair, which provides the guarantee of almost surely convergence {(following from martingale convergence theorem~\cite{grimmett2001probability})}.

{Objective (\ref{eq:obj 1})} is not the only choice for load flattening problems. A more intuitive approach is to directly minimize the variance of loads. This more intuitive formulation of load flattening problems has been widely employed by~\cite{clement2010impact,lopes_integration_2011,sortomme_coordinated_2011,mets_optimizing_2010}. Indeed, {minimizing load variance has been proved to be equivalent to (\ref{eq:obj 1}) in~\cite{gan2011optimal, abc}.} A general formulation for minimizing load variance can be written as

\begin{equation}
\begin{aligned}
\mathbf{Case\:2}: \, & \underset{p}{\text{min}}
& & \frac{1}{T}\sum_{t = 1}^{T} \Big\{\sum_{n=1}^{N} p(n,t) + P_{B}(t) \\
& & &\quad - \frac{1}{T}\sum_{\tau = 1}^{T}\Big[P_{B}(\tau) + \sum_{n=1}^{N}\hat{p}(n,\tau)\Big] \Big\}^{2},   \\
& \text{s.t.}
& & Eqn.\;(\ref{eq:constraint 1}),\;(\ref{eq:constraint 2}),
\label{formulation::loadvary}
\end{aligned}
\end{equation}
where $\hat{p}(n,\tau)$ denotes future charging loads, which can be obtained from prediction using historical data. This formulation can be viewed as to minimize the deviation from the average load (which is usually a constant). This is more intuitive as the optimal load profiles are designed to be as flat as possible.

However, {as shown in (\ref{formulation::loadvary}), future information (e.g., the base load $P_{B}(\tau)$, EV charging loads $\hat{p}(n,\tau)$, arrival and departure time of each EV, etc.) is required.} This is impractical when future information is hard to retrieve. Additionally, the performances of the corresponding algorithms are closely dependent on the accuracy of the predictions of future information.

{Instead of using offline approaches as in (\ref{eq:obj 1}) and (\ref{formulation::loadvary}),} two online heuristic approaches are separately proposed in~\cite{abc,gan2013realtime}. Li et al. utilize instant information to make a \textit{`myopic'} decision at each time step, and implement the allocation only for current step~\cite{abc}. Gan et al. solve the optimization problem formulated in~\cite{gan2011optimal} within each time step in~\cite{gan2013realtime} (the problem in~\cite{gan2011optimal} is a convex problem and can be rapidly solved). The variance of the {loads is minimized within the rest of the time horizon} at each time slot, then the allocation of charging power is implemented through a model predictive control (MPC) approach. The authors further prove that the expected variance of the aggregated load is linear proportional to the square of the prediction precision.
\subsubsection{Cost Minimization}
\label{subsubsec::cost}
Minimizing the overall  cost (energy generation cost and consumption cost) is another strong incentive for the aggregators to manipulate the schedule of EV charging. In term of energy generation cost, a instant generation cost curve has been provided in~\cite{masters_renewable_2013}. In term of energy consumption cost, a linear function has been used~\cite{ma_decentralized_2010} to model the electricity price with instant load. Thus the overall consumption cost is a quadratic function of an instant load. The total energy cost can be modelled as a cost function $\mathbb{C}(\cdot)$, with the aggregated EV charging loads and base loads as input. We first investigate a general formulation for generation cost as follows:

\begin{equation}
\begin{aligned}
\mathbf{Case\:3}: \; & \underset{p}{\text{min}}
& & \sum_{t = 1}^{T}\mathbb{C}\left (\sum_{n=1}^{N} p(n,t) + P_B(t)\right ), \\
& \text{s.t.}
& & Eqn.\;(\ref{eq:constraint 1}),\;(\ref{eq:constraint 2}).
\end{aligned}
\end{equation}

This formulation is employed by~\cite{ahn_optimal_2011} to minimize the combined cost of generation cost and tax charged due to generating carbon dioxide emissions. The generation cost is modelled as a piecewise linear function of produced power (similar to~\cite{masters_renewable_2013}). The authors further use conditional constraints to transfer the piece-wise linear function to a linear function,{ which can be solved with standard linear programming methods.}

In the case of reducing {EV charging cost}, the electricity price model provided in~\cite{ma_decentralized_2010} is utilized in~\cite{he_optimal_2012}. A general formulation of {EV charging cost minimization} is as follows:

\begin{equation}
\begin{aligned}
\mathbf{Case\:4}: \; & \underset{p}{\text{min}}
& & \sum_{t = 1}^{T}\Big\{ \alpha \Big[\sum_{n=1}^{N} p(n,t) + P_B(t)\Big] \\
& & & \quad + \beta\Big[\sum_{n=1}^{N} p(n,t) + P_B(t)\Big]^{2}\Big\},  \\
& \text{s.t.}
& & \! 0\!\! \leq E(n,0)\!\! +\!\! \sum_{\tau = 1}^{t} p(n,\tau)\Delta_{t} \leq E_{max}(n), \\
& & &  n = 1, \ldots, N,\quad t = 1, \ldots, T, \\
& & & \sum_{\tau = 1}^{t} p(n,\tau)\Delta_{t} \geq E(n,T) - E(n,0), \\
& & &  n = 1, \ldots, N, \quad t = 1, \ldots, T,\\
& & & Eqn.\;(\ref{eq:constraint 2}),
\label{formulation::case4}
\end{aligned}
\end{equation}
where the first constraint is the instant energy constraint, $E_{max}(n)$ is the maximal battery capacity of EV $n$. {The second constraint guarantees the all EVs can be charged to their predefined levels by the end of the time horizon.}
{He et al. \cite{he_optimal_2012}} solve (\ref{formulation::case4}) via a MPC-like approach, avoiding the need for a long term prediction of future information.

Mets et al. \cite{mets_optimizing_2010} compare the performance of global load flattening and local load flattening approaches. {While the former is for a global aggregator to coordinate EV charging loads,} the latter is described as local facilities that are only responsible for managing the loads within their own residential areas. {For both global load flattening and local load flattening problems, (\ref{formulation::case4}) has been used with global- and local-scale of aggregated information respectively.} The evaluation result in \cite{mets_optimizing_2010} demonstrates that, while both algorithms can effectively flatten the loads, the global load flattening algorithm has better performance.



\subsubsection{A General Formulation for Load Flattening Problems}
Load variance is not the only feature of the smart grid that will be impacted by integrating EVs. Other features including load factor and power losses will also be influenced if uncoordinated EV charging is prevalent. It has been proved in \cite{sortomme_coordinated_2011} that, given all the loads are connected to a single bus, problems of minimizing load variance, minimizing power losses and maximizing load factor form a \textit{`triangle equivalence'}. Although the assumption is not practical, simulations in \cite{sortomme_coordinated_2011} demonstrate a close approximation among these problems. Additionally, the authors have proved that the equivalence between minimizing load variance and maximizing load factor is topology independent. Moreover, as all of these problems are usually modelled as convex optimization problems, their corresponding algorithms can be considerably less computationally expensive (compared with the algorithm for minimizing power losses). Therefore, an effective algorithm solving the load flattening problem can not only approximately solve the other two problems, but also be more efficient for minimizing power losses problem especially.

To summarize optimization problems introduced in $\mathbf{Case\:1}$ to $\mathbf{Case\:4}$, a general formulation can be modelled as follows:
\begin{equation}
\begin{aligned}
\label{formulation::general}
& \underset{p}{\text{min}}
& & \sum_{t=1}^{T}\mathbb{F}\left (\sum_{n=1}^{N}p(n,t),S \right ), \\
& \text{s.t.}
& & Eqn.\;(\ref{eq:constraint 1}),\;(\ref{eq:constraint 2}).
\end{aligned}
\end{equation}

Various conditions can lead to different adaptions of (\ref{formulation::general}).
{First, $\mathbb{F}(\cdots)$ can be different objectives, including aggregated loads, aggregated energy or aggregated charging cost.} Secondly, when considering the availability of exogenous information, (1)~if the predicted information is available and sufficiently accurate, an off-line approach only require a one-time computation; (2)~if the exogenous information (e.g., the mobility information and demanded energy for reaching the destination of an EV) is difficult to predict, an online approach can utilize real-time obtained information to solve the problem. Thirdly, when given continuously controllable charging power, problems (\ref{formulation::general}) can be solved via classic optimization methods (e.g., linear programming, quadratic programming, etc.). However, if the charging power is assumed to be fixed (and even non-interrupted as in~\cite{gan_stochastic_2012}), the decision space is discrete. These problems can be coped via relaxation and approximation methods. We summarize the critical assumptions and the corresponding research works in Table.~\ref{Tab::Load}.


\begin{table*}
\begin{center}
\caption{Typical Research works categorized according to different conditions}
\label{Tab::Load}
\begin{tabular}{ccc}
\hline
Conditions & Terms & Articles \\ \hline
\multirow{4}{*}{Objective}                                                                 & Aggregated Loads                                                                     &     \cite{gan2011optimal,gan_stochastic_2012,ma_decentralized_2010}     \\
                                                                                           & \begin{tabular}[c]{@{}c@{}}Load Variance \\ (Power Losses, Load Factor)\end{tabular} &    \cite{gan2013realtime,clement2010impact,abc,mets_optimizing_2010,sortomme_coordinated_2011,Sundstroem2012Flexible}      \\
                                                                                           & Generation Cost                                                                      &     \cite{ahn_optimal_2011}     \\
                                                                                           & Charging Cost                                                                          &     \cite{he_optimal_2012,ma_decentralized_2010}     \\ \hline
\multirow{2}{*}{\begin{tabular}[c]{@{}c@{}}Charging Power \\ Controllability\end{tabular}} & Continuously Controllable                                                            &    \cite{gan2011optimal,ahn_optimal_2011,clement2010impact,gan2013realtime,he_optimal_2012,abc,ma_decentralized_2010,mets_optimizing_2010,sortomme_coordinated_2011,Sundstroem2012Flexible}      \\
                                                                                           & Fixed                                                                                &     \cite{gan_stochastic_2012}     \\ \hline
\multirow{2}{*}{Exogenous Information}                                                     & Predicted (Offline)                                                                  &    \cite{gan2011optimal,ahn_optimal_2011,clement2010impact,gan_stochastic_2012,mets_optimizing_2010,sortomme_coordinated_2011,Sundstroem2012Flexible}      \\
                                                                                           & real-time Collected (Online)                                                         &    \cite{gan2013realtime,he_optimal_2012,abc}       \\ \hline
\multirow{2}{*}{Solution}                                                                  & Centralized                                                                          &     \cite{clement2010impact,mets_optimizing_2010,sortomme_coordinated_2011,Sundstroem2012Flexible}       \\
                                                                                           & Decentralized                                                                        &     \cite{gan2011optimal,gan2013realtime,ahn_optimal_2011,gan_stochastic_2012,he_optimal_2012,abc,ma_decentralized_2010}     \\ \hline

\end{tabular}
\end{center}
\end{table*}

As we have demonstrated in Section. II, the smart grid interacts with EVs through bidirectional power flow. Until now we have only focused on the power flow from the smart grid to EV, investigating the influence on load variance, load factor and power losses by EV integration. The other direction, often referred to as V2G, depicts EVs providing auxiliary services to the smart grid, including voltage regulation, frequency regulation and spinning reserve etc.

Both directions of power flow can be managed through intelligent approaches as discussed in this section. For example, if we allow the minimum charging power $P_{min}(n,t)$ for EV $n$ to be negative, then the {load flattening methods studied in this section will evolve to peak shaving methods}. More specifically, each EV's rechargeable battery packs can be regarded as a distributed generator to reversely supply energy to the smart grid. Thus, the peak load can be shaved by these distributed rechargeable battery packs.

However, there is no doubt that V2G can induce many more interesting and challenging problems, especially for indirect control strategy design. This is because there are revenue incentives for both EV customers and EV aggregators. Meanwhile, the smart grid utilizes the revenue incentives to establish a more intelligent, efficient and reliable power system.
\subsection{Frequency Regulation}
\label{subsec:frequency}
There is another thread of research works handling the imbalance between power generation and demand through frequency regulation. Currently, it is generators, which is capable of providing MW-scale power, that are commonly utilized for frequency regulation \cite{Sekyung2010Development}. However, as we have previously discussed in both Section. \ref{subsubsec::coordination} and Section. \ref{subsec::flattening}, uncoordinated EV charging can bring severe impact to the balance of the smart grid. Fortunately, according to \cite{Kempton2005280}, PEVs are capable of making rapid response to frequency changes, thus regarded as suitable for providing frequency regulation service. One major obstacle for bring this to reality is the MW based minimum power capacity threshold \cite{kempton2005vehicle,Sekyung2010Development}. To conquer this obstacle, research works \cite{Sekyung2010Development,Sekyung2011Estimation,Chenye2012PEV,chenye2012vehicle} have depicted a scenario where an intermediate (referred to as an aggregator) aggregates small-scale power from EVs to provide frequency regulation service for the smart grid.

Frequency regulation and voltage regulation (discussed in following Section. \ref{subsec::voltage}) problems can be viewed as balancing active power and reactive power respectively \cite{Chenye2012PEV}. A coordination algorithm which jointly balances the supply and demand of both active and reactive power between PEVs and the smart grid has been developed in \cite{Chenye2012PEV}. The authors first propose a direct control approach, within which the central grid operator has control over all participated EVs and accurate energy status information. Then the central grid operator balance the active power via minimizing the deviation of real active power of each EV from the desired active power. This approach is much similar to \cite{clement2010impact,lopes_integration_2011,sortomme_coordinated_2011,mets_optimizing_2010} (introduced in Section. \ref{subsubsec::direct_control}), falling into the centralized coordination category which has rigid requirement as explained in Section. \ref{subsubsec::coordination}. Therefore, the authors further propose a price-based hybrid coordination algorithm which minimizes the cost of consuming active power. This approach is similar to \cite{mets_optimizing_2010,ahn_optimal_2011,he_optimal_2012} (discussed in Section. \ref{subsubsec::cost}), except that \cite{Chenye2012PEV} solves the optimization problem for each individual EV.

Han et al. propose several critical considerations for designing algorithms of EV-based frequency regulation in \cite{Sekyung2010Development}. The authors first differentiate providing regulation service and a normal charging operation from following aspect: (1) the payment for providing frequency regulation is based on the capacity of available power, while the expense for a charging operation is based on the amount of actual power dispatched. Then the authors further (2) restrict frequency regulation operations from interrupting charging operations, in that simultaneous charging and regulating may lead to serious generation oscillations (refer \cite{Sekyung2010Development} for detail explanations). Based on these considerations, a frequency regulation problem is formulated as follows:
\begin{equation}
\begin{aligned}
& \underset{C}{\text{min}}
& & \!\!\sum_{t=1}^{T} \Big[P_R(e(t),t) - C(t)P_R(e(t),t) - C(t)P_C(t)\Big]\\
& & & \quad -\lambda(e(T)-(E(T) - E(0)))^2, \\
& \text{s.t.}
& & e(t+1)=P_{max}C(t)+e(t), \; t = 1,\ldots, T, \\
& & &P_R(e(t),t) = K(t)e(t)+b(t), \; t = 1,\ldots, T, \\
\label{eq::fre_reg_1}
\end{aligned}
\end{equation}
where $P_{R}(e(t),t)$ and $P_{C}(t)$ are regulation price and charging price respectively, and $C(t)$ is the control sequence indicating when to initiate and cease a charging operation. Hence, $P_R(e(t),t) - C(t)P_R(e(t),t)$ is the revenue earned by providing regulation service, while $C(t)P_C(t)$ is the charging cost. $\lambda(e(T)-(E(T) - E(0)))^2$ is the penalty due to unsatisfied charging demand at deadline $T$. $P_{R}(e(t),t)$ is linearly dependent on $e(t)$, which is the state variable denoting the SOC at time $t$, $K(t)$ and $b(t)$ are coefficients varying over time. Note that an EV is charged by the maximal power $P_{max}$ once charging is initiated. The optimal control sequence is obtained by solving (\ref{eq::fre_reg_1}) with dynamic programming in \cite{Sekyung2010Development}.

Wu et al. depict a scenario where an aggregator possesses backup batteries and coordinate EV charging through a game theory approach \cite{chenye2012vehicle}. As stated in \cite{chenye2012vehicle}, backup batteries are intended to be rarely used, as frequent charging and discharging can lead to battery depreciation. Therefore, the objective is to minimize the usage of backup batteries within each time slot, while satisfying the demanded regulation capacity:
\begin{equation}
\begin{aligned}
& \underset{N_{c},N_{d}}{\text{min}}
& & \left |E(t) + R(t) + N_{d}(t) - N_{c}(t) - E_{ref}\right |\\
& \text{s.t.}
& & N_{d}(t) + N_{c}(t) \leq N, \; t = 1,\ldots, T, \\
& & & N_{d}(t) \geq 0,  N_{c}(t) \geq 0, \; t = 1,\ldots, T, \\
& & &\mathbf{d}(t)\cap\mathbf{c}(t) = \O , \; t = 1,\ldots, T, \\
\label{eq::game_frequency}
\end{aligned}
\end{equation}
where $E(t)$ is the storage status of backup batteries, $R(t)$ is the energy capacity requested by frequency regulation, $E_{ref}$ is the desired storage status that the aggregator intends to maintain. $\mathbf{d}(t)$ and $\mathbf{c}(t)$ denote the sets of all present EVs that choose to be discharged and charged respectively, $N_{d}$ and $N_{c}$ are the numbers of EVs belonging to $\mathbf{d}(t)$ and $\mathbf{c}(t)$.

The aggregator is assumed to coordinate EV charging operations by designing pricing policy for charging and discharging. The best response to the proposed smart pricing policy is a Nash Equilibrium, which maximizes the profits of all participated EVs. Additionally, the authors prove that the achieved Nash Equilibrium also solves (\ref{eq::game_frequency}).

Recall that the rigid requirements of a centralized coordination approach contain accurate signal acquisition. This is less practical when reliable communication of measurement data is not available or too expensive. Yang et al. tackle this problem in \cite{Hongming2013Application}, via proposing a distributed acquisition approach (based on consensus filtering) for obtaining consistent and accurate acquisition of frequency deviation signals for all participated EVs. With acquired frequency deviation, the authors propose a control theory based approach which integrates a dynamic EV model with frequency regulation.

\subsection{Voltage Regulation}
\label{subsec::voltage}
As frequency regulation aims at balancing the supply and demand of active power, voltage regulation is for maintaining the balance of reactive power. Traditionally, the central grid operator utilizes capacitors (e.g., switchable capacitors, shunt capacitors, static var compensators) for reactive power compensation \cite{Baran1989Optimal}. Recently, EVs have been shown to be capable of compensating reactive power via using their own on-board AC/DV inverters \cite{Wirasingha2008Plug}. Additionally, according to \cite{Kisacikoglu2010Examination}, EV customers can benefit from using EVs to provide voltage regulation without causing severe degradation to the batteries, which is usually more significant for frequency regulation when frequent charging and discharging is conducted. These properties have attracted many research interest \cite{Chenye2012PEV,Chenye2012PEVreactive,Mitsukuri2012Voltage,Chenye2014PQcontrol} of using EVs for realizing voltage regulation.

We first investigate \cite{Chenye2012PEV} as in Section. \ref{subsec:frequency}. In \cite{Chenye2012PEV}, reactive power $Q$ is treated similarly as active power $P$, i.e., a direct control and a price-based control are proposed. In both approaches, active power $P$ and reactive power $Q$ are coupled together by following constraint:
\begin{equation}
\begin{aligned}
P^{2} + Q^{2} \leq A_{max}^{2},
\end{aligned}
\label{eq::apparent_power}
\end{equation}
where $A_{max}$ is the maximal allowable apparent power $A = V\times I$, which is the product of grid voltage $V$ and charger current $I$. (\ref{eq::apparent_power}) is commonly used in subsequent research works investigated in the rest part of this section.

A hierachical voltage regulation strategy is proposed in \cite{Mitsukuri2012Voltage}. The authors first prove that the effect of tuning reactive power is superior to that of tuning active power for voltage regulation. Furthermore, the authors shows the superiority of tuning reactive power at nodes nearer to the target node. Based on this discovery, the proposed hierachical voltage regulation strategy triggers PQ control (reduce active power $P$ utilized for charging from maximum level to generate reactive power $Q$ according to (\ref{eq::apparent_power})) at nodes sequently according to the ascending order of their distances to the target node, at which voltage drop violation has been detected.

Another PQ control based voltage regulation algorithm is proposed in \cite{Chenye2014PQcontrol}. The objective in \cite{Chenye2014PQcontrol} consists of two parts: (1) a cost function $\mathbb{C}(P_{0})$ due to active power usage (as introduced in Section. \ref{subsubsec::cost}), where $P_{0}$ denotes the total active power drawn by the substation residing at the root of a distribution network, and (2) a voltage regulation function $f(\mathbf{V})=\underset{i}{\text{max}}\left | V_{i} - V_{0} \right |$, which indicates the maximum voltage deviation among $\mathbf{V} = [V_{1},\cdots,V_{N^B}]$ from the reference voltage $V_{0}$ at the substation. $\mathbf{V}$ is a set consisting of voltage $V_{i}$ at bus $i = 1,\ldots,N^{B}$. The objective is shown as follow:
\begin{equation}
\begin{aligned}
& \underset{N_{c},N_{d}}{\text{min}}
& & \mathbb{C}(P_{0}) + f(\mathbf{V})\\
& \text{s.t.}
& & I_{i} = \mathbb{T}_{I}(P_{i},Q_{i},V_{i})\\
& & & P_{i} = \mathbb{T}_{P}(P_{i-1},P^{B}_{i},I_{i},r_{i}) \\
& & & Q_{i} = \mathbb{T}_{Q}(Q_{i-1},Q^{B}_{i},I_{i},x_{i}) \\
& & & V_{i} = \mathbb{T}_{V}(P_{i-1},Q_{i-1},V_{i-1},I_{i},r_{i},x_{i}) \\
& & & E_{n,i,T_{n,i}}\!\! -\!\! E_{n,i,t}\! \leq p_{n,i}\Delta_{t} + (T_{n,i}\! - \!t)a_{n,i}\Delta_{t} \\
& & & p_{n,i}^{2}+q_{n,i}^{2} \leq a_{n,i}^{2}  \\
& & & 0 \leq P_{0} \leq P_{max},\\
& & & n = 1,\ldots, N^{i},\;i = 1,\ldots, N^{B},
\label{eq::pqcontrol}
\end{aligned}
\end{equation}
where $\mathbb{T}_{I}(\cdot)$, $\mathbb{T}_{P}(\cdot)$, $\mathbb{T}_{Q}(\cdot)$ and $\mathbb{T}_{V}(\cdot)$ are current, line active power flow, line reactive power flow and voltage transition functions for all buses, obtained by referring to the DistFlow model \cite{Baran1989Optimal}. $P_{i}$, $Q_{i}$, $P^{B}_{i}$, $Q^{B}_{i}$, $V_{i}$, $I_{i}$, $r_{i}$ and $x_{i}$ are the active power flow, reactive power flow, consumed active power, consumed reactive power, voltage, current, line resistance and reactance of bus $i$. The fifth constraint guarantees that EV $n$ residing at bus $i$ can be charged up to the desired level $E_{n,i,T_{n,i}}$ with active power $p_{n,i}$ by the charging dead line $T_{n,i}$, while not violating the maximal allowable apparent power $a_{n,i}$. $N^{i}$ and $N^{B}$ denotes the number of EVs charged at bus $i$ and the total number of buses. Problem (\ref{eq::pqcontrol}) is further relaxed to a convex optimization problem in \cite{Chenye2014PQcontrol}, by converting $I_{i} = \mathbb{T}_{I}(P_{i},Q_{i},V_{i})$ from an equality constraint to an inequality constraint $I_{i} \geq \mathbb{T}_{I}(P_{i},Q_{i},V_{i})$.

Wu et al. study voltage regulation problem when EVs charging stations are co-located with wind distributed generation units in \cite{Chenye2012PEVreactive}. Since wind turbines consume reactive power, the authors propose to utilize EVs and capacitors to compensate reactive power thus stabilizing the voltage. The problem is modeled as a Stackelberg game, which is solved via backward induction. The authors first obtain the Nash equilibrium in the second stage, given the pricing policy and demanded reactive power set by the wind distributed generation unit. This subgame perfect equilibrium guarantees that all EVs achieve the maximal profit. Then in the first stage, the payoff for the wind generation unit is maximized by tuning the aforementioned parameters passed to the second stage, which in turn help set the Nash equilibrium to the optimal value for the unit. Their optimal result leads to a case where all EVs provide the same amount of reactive power, which is either the extreme amount of reactive power that an EV can provide or consume, or the total compensation difference (the difference between demanded reactive power and the reactive power provided by capacitors) averaged by all EVs.


\subsection{Smart Grid Oriented Uncertainty}
Despite that most research works investigated in this section have rarely mentioned the case when exterior future information is not available, there do exist uncertain factors when developing coordination control algorithm for the smart grid operator. The main uncertainty for the smart grid operator originates from (1) load uncertainty and (2) regulation uncertainty. Both uncertainty factors are actually inherited from the same mobility uncertainty of an EV fleet, which consists of various arrival time and departing time, as well as the various storage status of EVs when arrived for charging. However, for the former, this mobility uncertainty leads to load uncertainty when EVs mainly draw energy from the smart grid. While for the latter, this uncertainty leads to regulation uncertainty when EVs provide regulation service to the smart grid.

Handling load uncertainty is essential for load flattening. For example, to minimize load variance or deviation, information about future load is necessary. According to \cite{clement2010impact,lopes_integration_2011,sortomme_coordinated_2011,mets_optimizing_2010,gan2013realtime}, the most common approach is to assume the availability of the forecast of future load, which can be obtained through learning historical data. On the other hand, regulation uncertainty has been considered in \cite{Sekyung2011Estimation}, which is an expansion to \cite{Sekyung2010Development} (introduced in Section. \ref{subsec:frequency}). In \cite{Sekyung2011Estimation}, EVs are categorized by their plug-in probabilities and capacities. EVs falling into the same category are modeled with a normal distribution, of which the mean and variance parameters vary from category to category. Since each category is modeled by a normal distribution, the summation of random variables of all categories still forms a normal distribution. Hence, the probability distribution of the available power capacity can be obtained, and further used for optimizing the profit of an aggregator.

\section{Aggregator Oriented EV Smart Charging} \label{sec:welfare}
In the previous section, even though many works employ decentralized methods, there still exists a centralized controller to update and publish intermediate information to all present EVs. This centralized aggregator can be regarded as the representative for the smart grid. However, as discussed in~\cite{mets_optimizing_2010}, the global control can be further broken down to multiple local aggregators, which manage loads within their local areas. These local aggregators can be regarded as the representative for local public facilities including substations, charging stations, parking lots and residential area.

The aggregators bridge {the power flow} between the smart grid and EV customers. More specifically, an aggregator can be responsible for maintaining a stable and reliable power system, while being responsible for satisfying EV customers' charging demand. {However, an aggregator has motivations of either saving the operation cost or earning more profits, therefore cannot be narrowed down to altruism.} In this section, we investigate how the aggregators pursue multiple goals including providing services to the smart grid and EV customers and manage their own profits. The smart interactions among all three involved parties helps make the future power system smart. Since we have investigated the part that the aggregator is solely responsible for the smart grid, here we focus on the interaction between the aggregator and EV customers.

\begin{figure}[t]
\centering
\epsfig{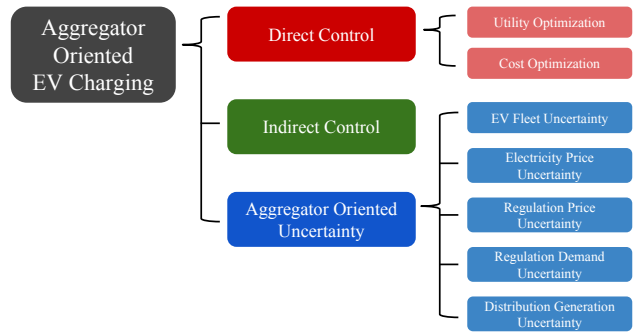}
\caption{Aggregator oriented EV charging.}
\label{fig::agg_perspect}
\end{figure}

In Section III, the service provided to an EV customer by an aggregator is a full charge guarantee at the charging deadline. However, in more practical situations, when an EV charging demand happens in a time limited situation, the satisfaction of a customer also includes how fast is the charging completed. Therefore, instant gain (e.g., instant charging power or instant SOC gain) is more desirable in these cases. Many research works have been conducted to maximize the overall satisfactions of all present EV customers. As discussed in Section III, the coordinated control can be either direct or indirect. Both control approaches will be investigated in this section, as shown in Fig. \ref{fig::agg_perspect}.

\subsection{Direct Coordinated Control}
EV charging problems can be studied in several ways through an algorithmic perspective. For example, an analogy between the distribution network with the Internet is proposed in~\cite{ardakanian_realtime_2012}. The authors propose an algorithm similar to that used to cope with a distributed computation system~\cite{low_optimization_1999}. Another thread of researches have regarded EVs charging problems as real-time resource scheduling problems~\cite{chen_large_2012,chen_deadline_2012,chen_iems_2012,subramanian_real-time_2012}. In these cases, each EV charging task is identified by their arrival time, deadline and processing time.

\subsubsection{Utility Optimization}
Research works falling into this category mainly focus on maximizing the overall welfare of EV customers. The overall welfare considered by an aggregator is interpreted as the aggregated satisfactions (utilities) of all EV customers. One model of the utility of an EV customer is a function of assigned instant charging power $p(n)$~\cite{ardakanian_distributed_2013}:

\begin{subequations}
\begin{align}
\mathbf{Case\:1}: \; & \underset{p}{\text{max}}
& & \sum_{n=1}^{N} \log(p(n)), \\
& \text{s.t.}
& & \sum_{n = 1}^{N_{l}} p(n) \leq P(l), \; l = 1, \ldots, L, \label{eq:constraint 3} \\
&&& P_{min}(n) \leq p(n)\leq P_{max}(n), \label{eq:constraint 4} \\
&&& n = 1, \ldots, N, \nonumber
\end{align}
\end{subequations}
where $l = 1, \ldots L$ denotes the distribution line and transformer where EV charging demands are aggregated. $P(l)$ is the capacity limit of the distribution line or transformer $l$.

The authors describe a future power grid distribution system equipped with measurement communication and control (MCC) nodes, which are capable of measuring and communicating in real-time. Thus, {real-time information of available capacity of the distribution system and EVs' SOC can be utilized to obtain optimal real-time charging power allocation decisions.}

Since (8) is a convex optimization problem, with constraints including coupled decision variables, it can be decomposed into a master problem and multiple subproblems. These problems are calculated by MCC nodes and EV smart chargers respectively. The MCC nodes communicate with all participating EVs simultaneously through broadcasting intermediate values to all present EVs. Then all EV adjusts the charging power immediately and simultaneously.

The utility function for an EV customer is a logarithmic function of assigned charging power $p(n)$. The choice of logarithmic function ensures that the optimal result also achieves proportional fairness. Since an EV customer' utility for the charging service is no longer a full charge guarantee at deadline for time limited cases, an online approach is needed to maximize the instant overall welfare. As the work in~\cite{ardakanian_distributed_2013} is for a static scenario for one snap shot of the system, this can be further expanded to a dynamic scenario~\cite{ardakanian_real-time_2014}. This extension is possible due to that the distributed power system is updated every milliseconds, much faster than real world events (e.g., EVs arrival and departure and the variation of base load). Therefore, each snap shot of this dynamic scenario can be viewed as static.
\subsubsection{Cost Optimization}
Another interpretation of individual welfare as the utility less the charging cost incurred is provided in~\cite{rahbari-asr_network_2013}. The formulation is as follows:
\begin{equation}
\begin{aligned}
\mathbf{Case\:2}: \; & \underset{p}{\text{max}}
& & \sum_{n=1}^{N} w_{n}(t)SOC_{n}^{t+1}(p(n,t)) \\
& & & \quad- \mathbb{C}_{n}^{t+1}(p(n,t)), \\
& \text{s.t.}
& & \sum_{n = 1}^{N} p(n,t) \leq P_{max,t}, \\
&&& Eqn.\;(\ref{eq:constraint 4}),
\label{wel::case2}
\end{aligned}
\end{equation}
where $w_{n}(t)$ is a positive coefficient to differentiate different EV customers. The SOC level for EV $n$ at next time slot is denoted by $SOC_{n}^{t+1}(\cdots)$, which is a function of current charging power $p(n,t)$. $\mathbb{C}_{n}^{t+1}(\cdots)$ is a linear cost function for $p(n,t)$. Since there is also coupled decision variables in the constraints, a decentralized solution can alleviate the computation burdens on the aggregator, while leading to a more robust and scalable problem. Therefore, Rahbari-Asr et al. employ an consensus algorithm in~\cite{rahbari-asr_network_2013} to solve this problem in a fully distributed way. The authors assume all EVs can {coordinate with their neighbours via peer-to-peer communication}. Furthermore, only next time slot SOC level is used as the gauge to evaluate an EV customer's utility in~\cite{rahbari-asr_cooperative_2014}. {The original objective of (\ref{wel::case2})} is approximated as a convex function for simplicity and solved with a consensus algorithm.

%
Real-time allocating charging power to multiple EVs can be viewed as real-time resource scheduling problem. Therefore, many classic real-time scheduling methods can be employed to solve EV charging power allocation problem. More specifically, the problem is to decide in real-time: (1)~which charging task is to be admitted and (2)~the charging schedule of each charging demand.
An admission control with greedy scheduling (TAGS) algorithm has been developed in~\cite{chen_deadline_2012,chen_large_2012,chen_optimizing_2013}. The authors combines an admission control strategy with earliest deadline first (EDF) algorithm. The admission decision is made by comparing the potential profits of admitting upcoming EVs with potential profits resulting from declining the requests. Once admitted, the scheduling is implemented in an EDF manner.

Subramanian et al. propose three heuristic real-time scheduling algorithms to tackle  EV charging power allocation problem with renewable energy integrated in~\cite{subramanian_real-time_2012}. The authors provide two heuristic algorithms based on EDF and least laxity first (LLF) real-time scheduling algorithms. The third algorithm employs a MPC approach using the forecast of future renewable generation.

\subsection{Indirect Coordinated Control}

In this part, we investigate research works using indirect control approaches for EV charging coordination. An indirect approach usually requires an aggregator to set a service price or revenue to attract EV customers to arrive. Meanwhile, the aggregator can earn revenue from the smart grid for completing the committed services, or pay penalty for the opposite case. For example, an aggregator can motivate EV customers to sell excess energy during peak hours in~\cite{tushar2012economics, tushar2014prioritizing}. The latter case is usually envisioned in a mature V2G market, where EVs are capable of providing auxiliary services to the smart grid \cite{chenye2012vehicle,Chenye2012PEV,Chenye2012PEVreactive,sortomme2012optimal,chunhua2013opportunities} (several research works \cite{chenye2012vehicle,Chenye2012PEV,Chenye2012PEVreactive} which propose price-based approaches for frequency regulation and voltage regulation have been investigated in Section. \ref{subsec:frequency} and Section. \ref{subsec::voltage}, hence they are omitted in this part).

There exists a trade-off between the service prices for EV customers and potential profits for the aggregators. For example, a much too high price can drive EV customers away, which would lead to less revenue earned from serving the customers. However, this can be necessary when the electricity price or loads are at peak. On the other hand, if the price is too low, although more customers would be attracted, there is a higher chance for this aggregator to overload the smart grid thus get penalized. {Therefore, this trade-off between charging services price and potential revenues of the aggregator needs to be tuned to achieve the maximal profits.}
\subsection{Aggregator Oriented Uncertainty}
An aggregator may confront many uncertainty factors when coordinating EV charging operations to support the smart gird. The uncertainty has its sources from EV customers, the smart grid, as well as distributed generation units if possessed by the aggregator. To summarize, when developing coordination control methods, following uncertainty factors need to be considered for improving the effectiveness:
\begin{enumerate}
\item EV fleet uncertainty (e.g., arriving time, departing time, energy status by arrival and actual power drawn);
\item Electricity price uncertainty (e.g. spot market price);
\item Regulation demand uncertainty (e.g., the amount of demanded capacity and regulation directions, i.e., regulation up and down);
\item Regulation price uncertainty;
\item Distribution generation uncertainty (e.g., power output of renewable generation units).
\end{enumerate}

These uncertainty factors have been partially tackled with in several research works \cite{Awami2012Coordinating,Vagropoulos2013Optimal,subramanian_real-time_2012}. EV fleet uncertainty has been considered in \cite{Vagropoulos2013Optimal}. In \cite{Vagropoulos2013Optimal}, the authors assume that behaviors of an EV fleet can be forecasted, since the statistical daily/weekyly commuting patterns of an EV fleet can be discovered by learning historical data. Similarly, information of both electricity price and regulation price are obtained via using a deterministic forecast in \cite{Awami2012Coordinating,Vagropoulos2013Optimal}. \cite{Vagropoulos2013Optimal} also considers regulation demand uncertainty, the estimate of regulation demand is obtained by statistically analysing historical data. Both \cite{Awami2012Coordinating} and \cite{Vagropoulos2013Optimal} employ a two-stage stochastic programming approach to solve the profit maximization problem \cite{Awami2012Coordinating} (cost minimization problem \cite{Vagropoulos2013Optimal}) for an aggregator. Additionally, a conditional-value-at-risk \cite{rockafellar2000optimization} term is appended in the objective of \cite{Vagropoulos2013Optimal}, in order to implement risk control. Distribution generation uncertainty is dealt with by utilizing deterministic forecast in \cite{subramanian_real-time_2012,Awami2012Coordinating}.

\section{Customer Oriented EV Smart Charging}
\label{sec:cost}

\begin{figure}[t]
\centering
\epsfig{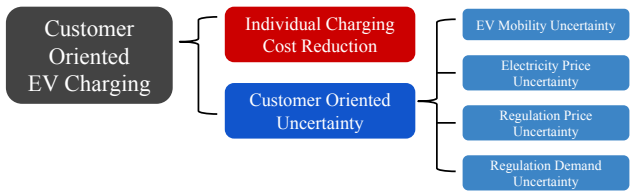}
\caption{Customer oriented EV charging.}
\label{fig::ev_perspect}
\end{figure}

We have investigated the cases in which an aggregator directly or indirectly control the charging for all EV customers in Section IV. However, a customer's top interest may not be to follow the control from the smart grid and satisfies the needs of the smart grid or the aggregator \cite{Parsons2014313}. Indeed, blindly following the requirement from the aggregator may not lead to the most beneficial result for an EV owner. For example, in a case of overnight charging, although an EV customer can have the EV fully charged, the charging profile designed by the aggregator may lead to more payment {when the charging services prices are varying in real-time.}

In this section, we focus on the perspective of an EV customer, who can earn benefits by practicing smart charging methods, as shown in Fig. \ref{fig::ev_perspect}. An ubiquitous smart charging management must respect an EV customer's freedom to minimize the charging cost. To achieve this objective, an EV owner needs future information including real-time service prices, regulation signals and regulation prices (if this EV customer participates in providing regulation services for the smart grid), {daily driving cycle and demanded energy to reach each intermediate points or the destination. The exogenous information can be obtained via prediction. However, when some information (e.g., regulation signals and real-time prices) is difficult to predict, an EV customer would confront a case of minimizing the expected charging cost within a specified time horizon.}
\subsection{Individual Charging Cost Reduction}
\label{subsec::indiv_ev_cost}
Research works falling into this category aim at minimizing the overall charging cost for an EV owner within a time horizon, while fulfilling the energy requests for arriving multiple intermediate stops and destination.

{Iversen et al. \cite{iversen_optimal_2014} minimize the combinational cost due to both an EV customer's charging cost and the penalty of maintaining less sufficient energy to reach each intermediate stops. The authors model this problem as a stochastic programming problem.} In this problem, the states are assumed to be random and the decision space is continuous. Additionally, the size of the decision space grows exponentially as more time steps are considered. However, to obtain a more accurate prediction of future information, a small time interval between each time step is desirable. The number of time steps can be even larger since a daily driving cycle is considered. {Therefore}, an exact solution to previous problem is hard to achieve. {Iversen et al. \cite{iversen_optimal_2014} discretize the state and decision space. This transforms the original problem into a discrete stochastic dynamic programming problem, which is solved through backward induction.
A similar model of individual EV customer's cost minimization problem is proposed in~\cite{rotering_optimal_2011}. The objective in~\cite{rotering_optimal_2011} is to minimize the charging cost for a PHEV during night time, with the guarantee of having the PHEV fully charged at the end of the horizon. The authors also assume that a PHEV owner can earn revenue by providing regulation service for the smart grid through an aggregator. Exogenous information including electricity price forward curves, regulation up and down curves, and energy required by the driving cycle is assumed to be available. Thus, this problem is formulated as a classical dynamic programming problem. Donadee et al.~\cite{donadee_stochastic_2014} further expand~\cite{rotering_optimal_2011} by integrating stochastic information {(hourly electricity prices, the hourly regulation service prices and hourly regulation signals) and formulate the problem as a stochastic dynamic programming problem. Similar to~\cite{iversen_optimal_2014}, this problem also suffers from randomness of the energy state space and continuousness of the decision space, thus need discretization of energy state space.}

{To summarize,} a general formulation of the individual charging cost reduction problem is as follows:
\begin{equation}
\begin{aligned}
& \underset{p}{\text{min}}
& & \mathbb{E}\left (\sum_{t=1}^{T} U(p(t),\mathbf{S}(t))\right ), \\
& \text{s.t.}
& &\sum_{t=1}^{T} p(t)\Delta_{t} \leq E(T) - E(0), \\
& & & P_{min}(t) \leq p(t)\leq P_{max}(t), \; t = 1,\ldots, T, \\
& & &\mathbf{S}(t) \sim \mathbb{T}_{t}(S(t-1)), \; t = 1,\ldots, T, \\
\end{aligned}
\end{equation}
where the bold $\mathbf{S}(t)$ denotes the set of random exogenous information, and $\mathbb{T}_{t}(\cdots)$ is a state transition function for cases involving a random process. This problem is usually dealt with discretization due to previous analysis of the problem size.

{Smart scheduling for individual EV customers' charging cost reduction have been rarely studied. This problem would be more prevalent as the penetration of EVs increases, thus need to be further explored. Especially when considering the nonlinear relation between charging time spent and the SOC obtained (Section. II), a smart scheduling for individual EV customers can provide charging schedules that have not only optimal charging cost, but also optimal charging time.}
\subsection{Customer Oriented Uncertainty}
Since an EV customer mainly interacts with an aggregator (which help bridge the interaction between this EV customer and the smart grid), this EV customer confronts less types of uncertainty. The main uncertainty studied by research works \cite{iversen_optimal_2014,donadee_stochastic_2014,Shi2011Real,mohsenian2015optimal} includes (1) EV mobility uncertainty, (2) electricity price uncertainty, (3) regulation price uncertainty and (4) regulation demand uncertainty. EV mobility uncertainty is mainly investigated in \cite{iversen_optimal_2014,mohsenian2015optimal}. Iversen et al. model an EV's driving pattern with a standard Markov model in \cite{iversen_optimal_2014}. While in \cite{mohsenian2015optimal}, the authors propose a real-time algorithm to minimize the expected charging cost given the conditional probability of departing time. Three other types of uncertainty are also modeled as a Markov random processes in \cite{donadee_stochastic_2014} (introduced in Section. \ref{subsec::indiv_ev_cost}). In \cite{Shi2011Real}, electricity price and regulation price are assumed to be unknown. The authors use a Markov chain to model the uncertain prices, and solve a Markov decision process problem which maximizes the profit of an EV customer, who can profit by choosing differet actions (charging, discharging and providing frequency regulation service). The authors further utilize an online Q-learning algorithm to learn the transition probabilities of electricity price and regulation price.

\section{Future Work}
\label{sec:future}

Many research efforts about EV charging have been developed based on assumptions listed in Table. I. While these assumptions can leave the problems well modelled and easily solved, they have also pointed out future research directions to reconsider these assumptions and solve more practical EV charging problems.

(1) Charging concerns due to battery properties: We have investigated many research works based on the assumption that battery SOC linearly depends on the charging time spent, given a constant charging power. However, as discussed in Section. II, the relationship between SOC and charging time is actually nonlinear. This overlook can lead to two folds of issues:
\begin{itemize}
\item Both the lifespan and available capacity of a battery pack can be shorten and degraded respectively, due to a long time of having the battery pack recharged and maintained near its full capacity.
\item Using a linear model for battery charging can incur extra charging cost and charging time spent for an EV customer, since the charging rate of a battery with higher SOC is slower than that of a battery with lower SOC. This problem can be severer when an EV customer needs to arrive at multiple intermediate stops and charge multiple times. Therefore, how to design a schedule for an EV customer to fulfil the daily driving cycle while spending minimal charging time or minimal cost is meaningful yet still unsolved.
\end{itemize}

(2) EV charging pattern estimate: Most of the research works investigated in this survey either assume the charging patterns of EVs (including when and how to charge EVs) are known a priori, or select distribution models (e.g. Gaussian distribution \cite{Lee2012Stochastic}). However, it is fundamental and crucial for an aggregator or ISO to have an effective and accurate estimate or forecast of the charging patterns of EVs. A potential solution can combine both statistic analysis and real-time feedback, which can be enabled by real-time EVs mobility data exchange between EVs and the smart grid.

(3) Develop more intelligent EV scheduling: From an individual EV customer's perspective, an intelligent integration of EVs includes not only smart scheduling of individual EV charging, following aspects also need attentions from both industry and academia:
\begin{itemize}
\item Range estimation, which is critical for an EV customer who suffers from range anxiety.
\item EV smart routing for discovering minimum energy consumed or minimum time needed, and exploring the trade-off between these two goals.
\item For an EV with hybrid energy sources (e.g. PHEV), it is rarely explored for the problem of finding a optimal strategy combining both EV routing and power splitting strategy, to further achieve a higher efficiency for both travelling time and consumed energy.
\end{itemize}

(4) Communication requirements: The communication networks is the essential foundation for realizing effective and efficient interactions among involved parties in the smart grid. According to \cite{doe2010}, the communication networks has requirements for communication latency, bandwidth and reliability as $10$-$100$ kbps, $2$-$5$ seconds and $99\%$-$99.99\%$ respectively. However, these requirements have not been comprehensively investigated, and will be more challenging to meet as the population of EVs increases in the future. Especially, severe communication delay can incur undesirable consequences for both EV customers and the smart grid:
\begin{itemize}
\item Communication delay can lead to incorrect routing and scheduling designs for EVs. For example, \cite{wang_mobility-aware_2014} has shown that the communication delay can result in extra energy cost as well as travelling time spent.
\item Communication delay can cause instability or even insecurity of the smart grid, as more and more research works considering implementing V2G to provide regulation service. For example, the smart grid can be overload if the system transformers or substations are already operating with their full capacities. To tackle this impact caused by communication delay, many well developed research methods regarding networked control system can be employed and even advanced.
\end{itemize}

(5) Communication security issues: Since both measurement data of EVs and coordination command from the smart grid are transmitted by the communication networks, following communication security issues also impose new threats on the smart interaction between EVs and the smart grid:
\begin{itemize}
\item Two-way communication between EVs and the smart grid will cause EV customers to transmit their private information (e.g. their locations, driving patterns, charging patterns and charging status, etc.) to third parties. Privacy leakage can be arisen if this information is exposed to unauthorized third parties.
\item One most common prerequisite for realizing coordinated EV charging is to grant the smart grid operator, or an aggregator, the ability to control EV charging operations. Therefore, EV charging need to be well protected from being manipulated by unauthorized third parties.
\item Besides the harmful consequences that an unauthorized third party can bring to EV customers (by manipulating their charging operations), this can also in turn bring severe threats to the stability and safety of the smart grid. There have been Internet-based load-altering attacks which overload the power grid by causing synchronous charging or discharging of PEVs \cite{mohsenianrad2011distributed,li2012securing,pasqualetti2013attack,hossain2012smart}.
\end{itemize}
\section{Conclusion}
\label{sec:conclusion}
This survey has reviewed research works focusing on the smart interactions among the smart grid, aggregators and EVs from an algorithmic perspective. Research works investigated are categorized according to their different standpoints, e.g., smart grid oriented EV smart charging, aggregator oriented EV smart charging and customer oriented EV smart charging. {For smart grid oriented EV smart charging, we have investigated the load flattening problems (approximately equivalent problems of minimizing power losses and increasing load factor) solved via optimization-based approaches, which has been broadly adopted.} In term of aggregator oriented EV smart charging, both direct control approach and indirect control approach have been addressed to maximize the overall satisfaction of all present EV customers. {As for customer oriented EV smart charging, we have studied stochastic optimization approaches to design the optimal schedule for minimizing individual EV customers' charging cost.

We have further pointed out several potential research directions based on the crucial and unique properties of {both EV rechargeable batteries and communication networks. Researches on these topics can help {accelerate the realization of the smart interaction between EVs and the smart grid.}

\bibliographystyle{IEEEtran}

\end{document}